\documentclass[
aps,
pre,
showpacs
]{revtex4-1}

\usepackage{braket}
\usepackage{delarray}
\usepackage{here}

\usepackage{txfonts}

\usepackage[dvipdfmx]{graphicx}
\usepackage{bm}
\usepackage{color}
\usepackage{ulem}

\newcommand{\be}{\begin{eqnarray}}
\newcommand{\ee}{\end{eqnarray}}
\newcommand{\ba}{\begin{array}}
\newcommand{\ea}{\end{array}}
\newcommand{\bmat}{\left(\begin{array}}
\newcommand{\emat}{\end{array}\right)}
\newcommand{\no}{\nonumber}

\newcommand{\Tr}{\mbox{Tr} }

\begin{document}
\title{Exact solution of free entropy for matrix-valued geometric Brownian motion with non-commutative matrices via replica method}
\author{Manaka Okuyama$^1$}
\author{Masayuki Ohzeki$^{1,2,3,4}$}
\affiliation{$^1$Graduate School of Information Sciences, Tohoku University, Sendai 980-8579, Japan}
\affiliation{$^2$International Research Frontier Initiative, Tokyo Institute of Technology, Tokyo 105-0023, Japan}
\affiliation{$^3$Department of Physics, Tokyo Institute of Technology, Tokyo 152-8551, Japan}
\affiliation{$^4$Sigma-i Co., Ltd., Tokyo 108-0075, Japan} %\\

\begin{abstract} 
Geometric Brownian motion (GBM) is a standard model in stochastic differential equations.
In this study, we consider a matrix-valued GBM with non-commutative matrices.
Introduction of non-commutative matrices into the matrix-valued GBM makes it difficult to obtain an exact solution because the existence of noise terms prevents diagonalization. However, we show that the replica method enables us to overcome this difficulty.
We map the trace of the time evolution operator of the matrix-valued GBM with non-commutative matrices into the partition function of the isotropic Lipkin-Meshkov-Glick model used in quantum spin systems.
Then, solving the eigenvalue problem of the isotropic Lipkin-Meshkov-Glick model, we obtain an analytical expression of the free entropy.
Numerical simulation is consistent with our analytical result.
Thus, our expression is the exact solution of the free entropy for the matrix-valued GBM with non-commutative matrices.
\end{abstract}
\date{\today}
\maketitle

%%%%%%%%%%%%%%%%%%%%%%%%%%%%%%%%%%%%%%%%%%%%%%%%%%%%%%%%%%%%%%%%%%%%%%%%%%%
%%%%%%%%%%%%%%%%%%%%%%%%%%%%%%%%%%%%%%%%%%%%%%%%%%%%%%%%%%%%%%%%%%%%%%%%%%%
\section{Introduction}
Stochastic differential equations, which are differential equations with randomness that varies with time, are used in various fields such as physics and mathematics.
Compared to the case of deterministic differential equations, it is more difficult to find an exact solution for a single trajectory in stochastic differential equations due to the effect of noise.
Geometric Brownian motion (GBM), which is one of the simplest stochastic differential equations, describes stock price movements in financial engineering~\cite{CZ,Duffie,Hull,HO,KS,KP} and is a solvable model in which an exact solution for a single trajectory can be obtained.
Thus, it is possible to obtain expectations for several quantities exactly.

A natural extension of solvable differential equations to multicomponent systems is a matrix form.
A GBM in matrix form~\cite{Hu} can be solved if all the matrices are commutative.
However, an exact solution for a single trajectory has not been obtained for the case where the matrices are non-commutative because the existence of the noise terms prevents diagonalization. This is a very different phenomenon from deterministic differential equations in matrix form.
The matrix-valued GBM with non-commutative matrices is a simple model, but finding its exact solution is a formidable problem.
 
In this study, we focus on the expectation of the logarithm of the trace of the time evolution operator, which corresponds to the free entropy in statistical physics, in the matrix-valued GBM with non-commutative matrices.
The free entropy is often an important quantity in stochastic differential equations and corresponds to the height of the surface in the Kardar-Parisi-Zhang equation~\cite{CDR,Dotsenko2,KPZ,BS} which is one of the most famous nonlinear stochastic partial differential equations in statistical physics.
Usually, it is difficult to attain the exact solution of the free entropy because an exact solution for a single trajectory cannot be obtained; however, we demonstrate that the free entropy for the matrix-valued GBM with non-commutative matrices can be obtained analytically.
The key ingredient of the derivation is the replica method~\cite{EA}, which was developed in the analysis of spin glass models in statistical physics~\cite{MGV,Nishimori,Dotsenko}.
Using the replica method, we map the trace of the time evolution operator of the matrix-valued GBM with non-commutative matrices into the partition function of the isotropic Lipkin-Meshkov-Glick model~\cite{LMG}, which is a mean-field model in quantum spin systems.
Then, through the isotropic Lipkin-Meshkov-Glick model, we obtain the exact solution of the free entropy for the matrix-valued GBM with non-commutative matrices.

Recently, an analytical solution of the Kardar-Parisi-Zhang equation was obtained by the replica method~\cite{CDR,Dotsenko2} and shown to coincide with the exact solution~\cite{SS,ACQ}. 
Our results can be regarded as applying their results to the simplest but non-trivial GBM system with discrete degrees of freedom.
This means that the replica method is also a very powerful tool for analyzing various stochastic differential equations in addition to the Kardar-Parisi-Zhang equation.

The organization of the paper is as follows.
In Sec. II, we define the matrix-valued GBM with non-commutative matrices.
In Sec. III, using the replica method, we map the trace of the time evolution operator of the matrix-valued GBM with non-commutative matrices into the partition function of the isotropic Lipkin-Meshkov-Glick model.
In Sec. IV, we consider the case of the even replica number and obtain the analytical expression of the free entropy for the matrix-valued GBM with non-commutative matrices.
In Sec. V, we consider the case of the odd replica number and obtain the analytical expression of a quantity different from the free entropy.
Finally, our conclusion is given in Sec. VI.

%%%%%%%%%%%%%%%%%%%%%%%%%%%%%%%%%%%%%%%%%%%%%%%%%%%%%%%%%%%%%%%%%
%%%%%%%%%%%%%%%%%%%%%%%%%%%%%%%%%%%%%%%%%%%%%%%%%%%%%%%%%%%%%%%%%
\section{Matrix-valued GBM with non-commutative matrices}
We consider the matrix-valued GBM with non-commutative matrices,
\be
d\hat{U}(t)&=& h \hat{\sigma}^z\hat{U}(t) dt +\sqrt{q_x}\hat{\sigma}^x\hat{U}(t)  \circ dW_{x}(t)+\sqrt{q_y}\hat{\sigma}^y\hat{U}(t)  \circ dW_{y}(t)+\sqrt{q_z}\hat{\sigma}^z\hat{U}(t)  \circ dW_{z}(t)  ,\label{GBM-Stra}
\no\\
\\
\hat{U}(0)&=&
\left(\begin{array}{cc}
1 & 0 \\
 0 & 1
\end{array}\right),
\\
&&0\le t \le \beta ,
\ee
where $h$, $q_x$, $q_y$, and $q_z$ are time-independent constants; $\hat{\sigma}^a$ $(a=x,y,z)$ is the Pauli matrix; and $W_{x}(t)$, $W_{y}(t)$, and $W_{z}(t)$ denote standard  Brownian motions.
The symbol ``$\circ$" denotes the Stratonovich interpretation. 
The equivalent Ito process is given by
\be
d\hat{U}(t)&=&h \hat{\sigma}^z\hat{U}(t) dt +\sqrt{q_x}\hat{\sigma}^x\hat{U}(t)  \bullet dW_{x}(t)+\sqrt{q_y}\hat{\sigma}^y\hat{U}(t)  \bullet dW_{y}(t)+\sqrt{q_z}\hat{\sigma}^z\hat{U}(t)  \bullet dW_{z}(t)  
 +\frac{q_x + q_y + q_z}{2}  \hat{U}(t)   dt , \label{Ito}
\ee
where the symbol ``$\bullet$" denotes the Ito interpretation. 

Here, from the analogy of the free entropy in statistical physics, we are interested in the expectation of the logarithm of the trace of the time evolution operator,
\be
\mathbb{E} \left[ \log\left( \Tr \left(\hat{U}(\beta)\right) \right) \right],
\ee
and also call this the free entropy. 

When $q_x=q_y=0$, all the matrices commute with each other, and we can obtain the exact solution for a single trajectory,
\be
\hat{U}(t)&=&
\left(\begin{array}{cc}
\exp\left(th+\sqrt{q_z}W_z(t)\right) & 0 \\
 0 & \exp\left(-t h-\sqrt{q_z}W_z(t)\right)
\end{array}\right).
\ee
Then, we immediately obtain the exact solution of the free entropy in the following manner:
\be
\mathbb{E} \left[ \log\left( \Tr \left(\hat{U}(\beta)\right) \right) \right] &=& \int_{-\infty}^{\infty} \frac{dW_z}{\sqrt{2\pi \beta}} e^{-\frac{W_z^2}{2\beta}} \log \left(2\cosh\left(\beta h+\sqrt{q_z}W_z\right) \right)
\no\\
&=& \int_{-\infty}^{\infty} \frac{du}{\sqrt{2\pi }} e^{-\frac{u^2}{2}} \log \left(2\cosh\left(\sqrt{\beta q_z}u +\beta h \right) \right). \label{exact-trivial}
\ee
However, for $q_x\neq0$ and $q_y\neq0$, the exact solution for a single trajectory has not been obtained because we cannot diagonalize the stochastic differential equation due to the noise terms $dW_{x}(t)$ and $dW_{y}(t)$.
Thus, it is also difficult to obtain the exact solution of the free entropy.

In this study, we focus on the case of $q_x=q_y=q\neq0$.
Using the replica method, we show that even though no exact solution for a single trajectory can be derived, we can obtain the exact solution of the free entropy.
The procedure of the replica method is as follows.
First, we obtain an analytical expression of $\mathbb{E} \left[ \left( \Tr \left(\hat{U}(\beta)\right) \right)^n\right]$ for a natural number $n$.
Next, using analytic continuation of $n$ to real numbers and taking the limit $n\rightarrow0$, we obtain an analytical expression of the free entropy as follows:
\be
\lim_{n\rightarrow0} \frac{ \mathbb{E} \left[ \left( \Tr \left(\hat{U}(\beta)\right) \right)^n\right]-1}{n} =\mathbb{E} \left[ \log\left( \Tr \left(\hat{U}(\beta)\right) \right) \right].
\ee
Finally, in order to verify the correctness of the attained analytical expression, we compare it with a numerical simulation.

%%%%%%%%%%%%%%%%%%%%%%%%%%%%%%%%%%%%%%%%%%%%%%%%%%%%
%%%%%%%%%%%%%%%%%%%%%%%%%%%%%%%%%%%%%%%%%%%%%%%%%%%%
\section{Mapping into isotropic Lipkin-Meshkov-Glick model}
From the definition of Eq. (\ref{GBM-Stra}), the expectation of the trace of the time evolution operator can be represented as
\be
\mathbb{E} \left[  \Tr \left(\hat{U}(\beta) \right) \right]&=& \lim_{M\rightarrow\infty}\mathbb{E} \left[ \Tr \left(\prod_{t=1}^M e^{ \frac{\sqrt{\beta q_{}}}{\sqrt{M}}  J_{x,t} \hat{\sigma}^x }e^{ \frac{\sqrt{\beta q_{}}}{\sqrt{M}}  J_{y,t} \hat{\sigma}^y }e^{ \frac{\sqrt{\beta q_{z}}}{\sqrt{M}}  J_{z,t} \hat{\sigma}_z +\frac{\beta}{M}  h \hat{\sigma}^z} \right) \right] ,
\ee
where $J_{x,t}$, $J_{y,t}$, and $J_{z,t}$ follow Gaussian distributions with mean 0 and variance 1.
Then, by calculating the averages of $J_{x,t}$, $J_{y,t}$, and $J_{z,t}$, we arrive at
\be
\mathbb{E} \left[ \left( \Tr \hat{U}(\beta) \right)^n\right]&=&\lim_{M\rightarrow\infty}\Tr \left( \prod_{t=1}^Me^{ \frac{\beta}{M}  \frac{q}{2} \left(\sum_{\alpha=1}^n  \hat{\sigma}_{\alpha}^x \right)^2} e^{ \frac{\beta}{M} \frac{q}{2} \left(\sum_{\alpha=1}^n  \hat{\sigma}_{\alpha}^y \right)^2 }e^{\frac{\beta}{M} \frac{q_z}{2} \left(\sum_{\alpha=1}^n  \hat{\sigma}_{\alpha}^z \right)^2
+\frac{\beta}{M} h \sum_{\alpha=1}^n  \hat{\sigma}_{\alpha}^z}\right)
\no\\
&=&\Tr \left( e^{ -\beta \hat{H}_{LMG}(n)}\right) \label{LMG-partition},
\ee
where $\hat{H}_{LMG}(n)$ is the isotropic Lipkin-Meshkov-Glick model~\cite{LMG},
\be
\hat{H}_{LMG}(n)&=&  - \frac{q}{2} \left(\sum_{\alpha=1}^n  \hat{\sigma}_{\alpha}^x \right)^2 - \frac{q}{2} \left(\sum_{\alpha=1}^n  \hat{\sigma}_{\alpha}^y \right)^2- \frac{q_z}{2} \left(\sum_{\alpha=1}^n  \hat{\sigma}_{\alpha}^z \right)^2
-h \sum_{\alpha=1}^n  \hat{\sigma}_{\alpha}^z .
\ee
Thus, the problem is reduced to obtaining the partition function of the isotropic Lipkin-Meshkov-Glick model. 

In the following, we briefly summarize the eigenvalue problem of the isotropic Lipkin-Meshkov-Glick model~\cite{CFLRT}.
Introducing the total spin operator $\hat{S}^a=\sum_{i=1}^n \hat{\sigma}_i^a/2$, $\hat{H}_{LMG}(n)$ can be written as
\be
\hat{H}_{LMG}(n)&=&  - 2q\left( {\bf \hat{S}}^2 - (\hat{S}^z)^2 \right)- 2q_z (\hat{S}^z)^2-2 h \hat{S}^z.
\ee
We note that $\hat{H}_{LMG}(n)$, ${\bf \hat{S}}^2$, and $\hat{S}^z$ commute with each other,
\be
\left[\hat{H}_{LMG}(n) , \hat{S}^z \right]&=&0,
\\
\left[\hat{H}_{LMG}(n), {\bf \hat{S}}^2 \right]&=&0,
\\
\left[{\bf \hat{S}}^2 , \hat{S}^z \right]&=&0.
\ee
The eigenvectors of ${\bf \hat{S}}^2$ and $\hat{S}^z$ satisfy
\be
{\bf \hat{S}}^2 |S,M \rangle&=&S(S+1)|S,M \rangle ,
\\
\hat{S}^z |S,M \rangle&=&M|S,M \rangle ,
\ee
where $S$ and $M$ take the forms 
\be
S&=& \frac{n}{2}-K \   \left(K=0,1,2,\cdots, \left\lfloor \frac{n}{2} \right\rfloor \right) , \label{eigenvalue-S}
\\
M&=&-S,-S+1,\cdots, S-1, S ,
\ee
where $\left\lfloor \cdot \right\rfloor$ is the floor function.
Then, the eigenvalues of $\hat{H}_{LMG}(n)$ are given by
\be
E_{LMG}(S,M)&=&  - 2q\left( S(S+1) - M^2 \right)- 2q_z M^2-2 h M.
\ee
In addition, the energy eigenvalue $E_{LMG}(S,M)$ in sector $S$ is degenerate in the following numbers.
\be
 {}_n C _K \frac{n+1-2K}{n+1-K}= {}_n C _{\frac{n}{2}-S} \frac{1+2S}{1+S+\frac{n}{2}} .
\ee
In summary, the partition function of the isotropic Lipkin-Meshkov-Glick model for even $n$ is given by 
\be
\Tr\left( e^{-\beta\hat{H}_{LMG}(n)  }\right)&=&\sum_{S,M}e^{-\beta E_{LMG}(S,M)  } 
\no\\
&=&\sum_{S=0,1,\cdots,n/2}{}_n C _{\frac{n}{2}-S} \frac{1+2S}{1+S+\frac{n}{2}} e^{ 2\beta q S(S+1)}  \sum_{M=-S}^S  e^{ \beta\left(   2(q_z-q) M^2+2 h M \right)}.
\label{partition-even}
\ee
On the other hand, the partition function for odd $n$ is as follows:
\be
\Tr\left( e^{-\beta\hat{H}_{LMG}(n)  }\right)&=&\sum_{S=1/2,3/2,\cdots,n/2}^{} {}_n C _{\frac{n}{2}-S} \frac{1+2S}{1+S+\frac{n}{2}} e^{ 2\beta q S(S+1)}  \sum_{M=-S}^S  e^{ \beta\left(   2(q_z-q) M^2+2 h M \right)} .\label{partition-odd}
\ee

We note that the partition function depends on the parity of the number of replicas due to the property of $S$ in Eq (\ref{eigenvalue-S}).
This property must be treated carefully in proceeding with calculations.
For even $n$, the sum variable takes $S=0,1,\cdots n/2$ in Eq. (\ref{partition-even}), and it is possible to take the limit $n\to0$ naturally.
Thus, the free entropy is derived from the partition function of the even replica number with the limit $n\to0$.
On the other hand, for odd $n$, the sum variable takes $S=1/2,3/2,\cdots,n/2$ in Eq. (\ref{partition-odd}), and the operation of taking the limit $n\to0$ is not justified.
This means that the partition function of the odd replica number is related to other quantities different from the free entropy.
We will discuss the case of the odd number in more detail in Sec. \ref{n-odd}.

In the following, we first consider the case where $n$ is an even number.

%%%%%%%%%%%%%%%%%%%%%%%%%%%%%%%%%%%%%%%%%%%%%%%%%%%%
\section{Exact solution of free entropy: even $n$}

%%%%%%%%%%%%%%%%%%%%%%%%%%%%%%%%%%%%%%%%%%%%%%%%%%%%
%%%%%%%%%%%%%%%%%%%%%%%%%%%%%%%%%%%%%%%%%%%%%%%%%%%%
\subsection{Case with $q_z\ge q \ge 0$}
In order to calculate the sum of the partition function, we need to reduce the second-order term of $M$ and $S$ of the exponential function to the first-order term.
Using the Hubbard-Stratonovich transformation, we obtain the following relations:
\be
e^{ \beta\left(   2(q_z-q) M^2+2 h M \right)}
&=&  \int Du e^{ A_1 M },
\ee
and
\be
e^{2\beta q S(S+1)} 
&=&\int Dw e^{ A_2 S}, \label{hubbst-S}
\ee
where $\int Du\equiv \int_{-\infty}^\infty du e^{-u^2/2}/ \sqrt{2\pi}$, $A_1= 2\left(\sqrt{\beta(q_z-q)} u+\beta h \right) $, $\int Dw\equiv \int_{-\infty}^\infty dw e^{-w^2/2}/ \sqrt{2\pi}$, and $A_2=2\sqrt{\beta q}w+ 2\beta q$.

Then, we can take the sum on $M$ in Eq. (\ref{partition-even}) as follows:
\be
\sum_{M=-S}^S e^{ \beta\left(   2(q_z-q) M^2+2 h M \right)}
&=& \int Du\sum_{M=-S}^S e^{A_1 M}
\no\\
&=& \int Du\frac{e^{ - A_1 S  }-e^{  A_1 (S+1) } }{1-e^{  A_1 }}. \label{M-sum-after}
\ee
Using Eqs. (\ref{partition-even}), (\ref{hubbst-S}), and (\ref{M-sum-after}), we can rewrite the partition function as
\be
&&\Tr\left( e^{-\beta\hat{H}_{LMG}(n)  }\right)
\no\\
 &=& \int Dw \int Du \sum_{S=0}^{n/2}{}_n C _{\frac{n}{2}-S} \frac{1+2S}{1+S+\frac{n}{2}} e^{ A_2 S} \frac{e^{ - A_1 S  }-e^{  A_1 (S+1) } }{1-e^{  A_1 }}
\no\\
&=& \int Dw \int Du   \sum_{K=0}^{n/2 }  {}_n C _K \frac{n+1-2K}{n+1-K} e^{A_2(n/2-K)}\frac{\sinh(A_1(n/2-K+1/2))}{\sinh(A_1/2)} . \label{partition-1}
\ee
Furthermore, we can express the summation in Eq. (\ref{partition-1}) as
\be
&&\sum_{K=0}^{n/2 }  {}_n C _K \frac{n+1-2K}{n+1-K} e^{A_2(n/2-K)}\frac{\sinh(A_1(n/2-K+1/2))}{\sinh(A_1/2)}
\no\\
&=&\frac{e^{-2A_2}}{(-1+e^{A_1})}
\no\\
&& \left\{  e^{A_2-\frac{A_1 n}{2}+\frac{A_2 n}{2}}\left(-e^{A_1 n}(1+e^{-A_1-A_2})^n + e^{A_1+A_2+A_1 n} (1+e^{-A_1-A_2})^n + e^{A_1}(1+e^{A_1-A_2})^n-e^{A_2}(1+e^{A_1-A_2})^n\right) \right.
\no\\
&&\left.+ \frac{2 e^{A_2}}{n} {}_n C_{1+\frac{n}{2}} \left({}_2 F_{1}\left[1,-\frac{n}{2}, 2+\frac{n}{2},-e^{-A_1-A_2} \right] -e^{A_1} {}_2 F_{1}\left[1,-\frac{n}{2}, 2+\frac{n}{2},-e^{A_1-A_2} \right] \right) \right.
\no\\
&& \left. + \frac{4 e^{-A_1}}{n-2} {}_n C_{2+\frac{n}{2}} \left({}_2 F_{1}\left[2,1-\frac{n}{2}, 3+\frac{n}{2},-e^{-A_1-A_2} \right] -e^{3A_1} {}_2 F_{1}\left[2,1-\frac{n}{2}, 3+\frac{n}{2},-e^{A_1-A_2} \right] \right)  \right\},
\ee
where ${}_2 F_{1}\left[a,b, c, z \right]$ is the Gaussian hypergeometric function defined as
\be
{}_2 F_{1}\left[a,b, c, z \right]&=&\sum_{k=0}^{\infty} \frac{(a)_k (b)_k}{(c)_k k!}z^k,
\ee
where $(a)_k$ is the Pochhammer symbol.
Then, using an analytic continuation, we can realize the $n\rightarrow 0$ limit in the following manner:
\be
&&\lim_{n\rightarrow0} \left( \sum_{K=0}^{n/2 }  {}_n C _K \frac{n+1-2K}{n+1-K} e^{A_2(n/2-K)}\frac{\sinh(A_1(n/2-K+1/2))}{\sinh(A_1/2)} -1\right)/n
\no\\
&=&\frac{e^{-2A_2}}{2(-1+e^{A_1})} \left(1-A_1 - A_2 -e^{A_1} - A_1e^{A_1} + A_2 e^{A_1} + A_1 e^{A_2} -A_2 e^{A_2} +(A_1 + A_2) e^{A_1+A_2} \right.
\no\\
&& \left. -2 \log(1+e^{-A_1-A_2}) +2e^{A_1} \log(1+e^{A_1-A_2}) +{}_2 F_{1}^{(0,0,1,0)} \left[1,0,2, -e^{-A_1-A_2}\right] \right.
\no\\
&& \left. -e^{A_1}  {}_2 F_{1}^{(0,0,1,0)} \left[1,0,2, -e^{A_1-A_2}\right]-{}_2 F_{1}^{(0,1,0,0)} \left[1,0,2, -e^{-A_1-A_2}\right] +e^{A_1}{}_2 F_{1}^{(0,1,0,0)} \left[1,0,2, -e^{A_1-A_2}\right]\right),\label{sum-n->0}
\no\\
\ee
where ${}_2 F_{1}^{(0,0,1,0)} \left[x,y,z,w\right]$ means
\be
{}_2 F_{1}^{(0,0,1,0)} \left[x,y,z,w\right]&=&\frac{\partial}{\partial z} {}_2 F_{1}^{} \left[x,y,z,w\right].
\ee
Moreover, using the relations
\be
{}_2 F_{1}^{(0,0,1,0)}(1,0,2,y) &=&0,
\\
{}_2 F_{1}^{(0,1,0,0)}(1,0,2,y) &=&1+(-1+1/y)\log(1-y),
\ee
we can rewrite Eq. (\ref{sum-n->0}) as
\be
&&\lim_{n\rightarrow0} \left( \sum_{K=0}^{n/2 }  {}_n C _K \frac{n+1-2K}{n+1-K} e^{A_2(n/2-K)}\frac{\sinh(A_1(n/2-K+1/2))}{\sinh(A_1/2)} -1\right)/n
\no\\
&=&\frac{e^{-A_2}}{2(-1+e^{A_1})} \left(A_1(1+e^{A_1})(-1+e^{A_2})+ (-1+e^{A_1+A_2})\log(e^{-A_1}+e^{A_2}) +(e^{A_1}-e^{A_2})\log(e^{A_1}+e^{A_2})  \right)
\no\\
&=&\frac{1}{2(-1+e^{A_1})} \left(\{e^{A_1-A_2}-1 \}\log(1+e^{A_2-A_1}) +\{ e^{A_1}-e^{-A_2} \}\log(1+e^{A_2+A_1}) \right) . \label{sum2-n->0}
\ee
Finally, substituting Eq. (\ref{sum2-n->0}) into Eq. (\ref{partition-1}), we obtain an analytical expression of the free entropy as follows:
\be
&&\mathbb{E} \left[ \log\left( \Tr \hat{U}(\beta) \right) \right] 
\no\\
&=&\lim_{n\rightarrow0} \frac{\Tr\left(e^{-\beta \hat{H}_{LMG}(n)}\right)-1}{n}
\no\\
&=&\int Dw \int Du \frac{\left(\{e^{A_1-A_2}-1 \}\log(1+e^{A_2-A_1}) +\{ e^{A_1}-e^{-A_2} \}\log(1+e^{A_2+A_1}) \right)}{2(-1+e^{A_1})} .\label{exact-solution1}
\ee
When $q=0$ ($A_2=0$), our solution expressed in Eq. (\ref{exact-solution1}) coincides certainly with the existing exact solution, Eq. (\ref{exact-trivial}).
In addition, for $q=0$, the variable $w$ in Eq. (\ref{exact-solution1}) corresponds to the stochastic fluctuation of $W_{z}(t)$ in Eq. (\ref{Ito}).
Thus, for $q\neq0$, it is considered that the variable $u$ in Eq. (\ref{exact-solution1}) represents the stochastic fluctuations of both $W_{x}(t)$ and $W_{y}(t)$ in Eq. (\ref{Ito}).
It is important to note that there is only one degree of freedom of integration for two stochastic fluctuations $W_{x}(t)$ and $W_{y}(t)$.
This is because we consider the symmetric stochastic fluctuations, $q_x=q_y=q$.

%%%%%%%%%%%%%%%%%%%%%%%%%%%%%%%%%%%%%%%%%%%%%%%%%%%%
%%%%%%%%%%%%%%%%%%%%%%%%%%%%%%%%%%%%%%%%%%%%%%%%%%%%
\subsection{Case with $q \ge q_z\ge0$}
Next, we consider the case of $q \ge q_z \ge0$.
Using the Hubbard-Stratonovich transformation, we obtain the following relation:
\be
e^{ \beta\left( 2(q_z-q) M^2+2 h M \right)}
&=&  \int Du e^{ A_3 M },
\ee
where $A_3= 2\left(i\sqrt{\beta(q-q_z)} u+\beta h \right)$.
We note that $A_3$ is a complex number.
Then, following the same procedure as before, we obtain an analytical solution as follows: 
\be
&&\mathbb{E} \left[ \log\left( \Tr \hat{U}(\beta) \right) \right] 
\no\\
&=&\int Dw \int Du \Re\left( \frac{\left(\{e^{A_3-A_2}-1 \}\log(1+e^{A_2-A_3}) +\{ e^{A_3}-e^{-A_2} \}\log(1+e^{A_2+A_3}) \right)}{2(-1+e^{A_3})}  \right). \label{exact-solution2}
\ee

%%%%%%%%%%%%%%%%%%%%%%%%%%%%%%%%%%%%%%%%%%%%%%%%%%%%
%%%%%%%%%%%%%%%%%%%%%%%%%%%%%%%%%%%%%%%%%%%%%%%%%%%%
\subsection{Numerical verification of free entropy}
\begin{figure}[H]
\begin{tabular}{cc}
\begin{minipage}[t]{0.45\hsize}
\centering
\includegraphics[scale=0.53]{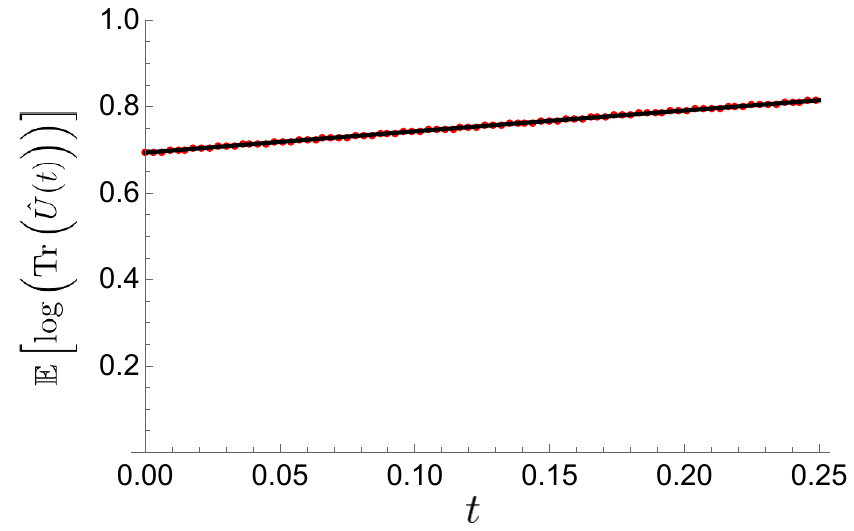}
\end{minipage} &
%%%%%%%%%%%%%%%%%%%%%%%%%%%%%%%%%%%%%%%%%%%%%%%%%%%%%%%%%%%%%%%%%
\begin{minipage}[t]{0.45\hsize}
\centering
\includegraphics[scale=0.53]{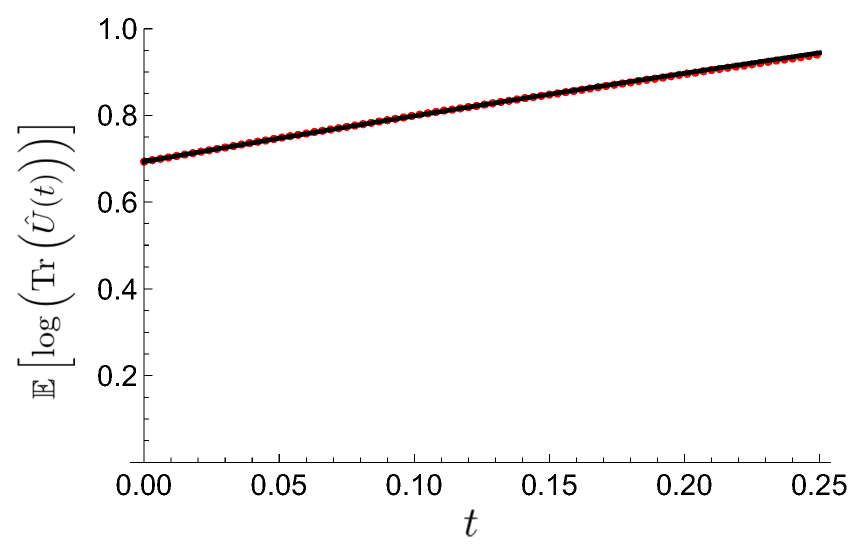}
\end{minipage}
\end{tabular}
\caption{Time dependence of the free entropy for the matrix-valued GBM with non-commutative matrices. 
The horizontal axis denotes the time. The vertical axis denotes the free entropy.
Left panel: For $h=1/5$, $q=1/5$, and $q_z=1/2$, the thick black line and the red circles show Eq. (\ref{exact-solution1}) and the numerical simulation of Eq. (4), respectively.
Right pane: For $h=1/10$, $q=1$ and $q_z=1/5$, the thick black line and the red circles show Eq. (\ref{exact-solution2}) and the numerical simulation of Eq. (4), respectively.}
\label{fig1}
\end{figure}

We have obtained the analytical expressions of the free entropy for the matrix-valued GBM with non-commutative matrices by the replica method.
However, we used the analytic continuation $n\rightarrow0$, which is not mathematically justified.
Thus, we need to pay attention to the validity of the obtained analytical expressions.
In order to verify our expressions, we compared Eqs. (\ref{exact-solution1}) and (\ref{exact-solution2}) with a numerical simulation of Eq. (\ref{Ito}) in Fig \ref{fig1}.
We chose the final time $\beta = 0.25$, the sample number $N=10000$, and the step size $\Delta t=1/10000$.
The error bars are given by the standard error but they are too small to be seen in Fig. \ref{fig1}.
Figure \ref{fig1} shows that our expression, Eq. (\ref{exact-solution1}), is precisely consistent with the numerical simulation of Eq. (\ref{Ito}).
Thus, this verifies that we have obtained the exact solution of the free entropy.

%%%%%%%%%%%%%%%%%%%%%%%%%%%%%%%%%%%%%%%%%%%%%%%%%%%%%%%%%%%%%%%%%

%%%%%%%%%%%%%%%%%%%%%%%%%%%%%%%%%%%%%%%%%%%%%%%%%%%%
%%%%%%%%%%%%%%%%%%%%%%%%%%%%%%%%%%%%%%%%%%%%%%%%%%%%
\section{Exact solution of $\mathbb{E} \left[ \Tr \left(\hat{U}(t)\right)\log\left( \Tr \left(\hat{U}(t)\right) \right) \right]$: odd $n$} \label{n-odd}
%%%%%%%%%%%%%%%%%%%%%%%%%%%%%%%%%%%%%%%%%%%%%%%%%%%%
%%%%%%%%%%%%%%%%%%%%%%%%%%%%%%%%%%%%%%%%%%%%%%%%%%%%
We have considered the case where $n$ is an even number.
In the following, we consider the case where $n$ is an odd number.
Then, we show that we can obtain an exact solution of $\mathbb{E} \left[ \Tr \left(\hat{U}(t)\right)\log\left( \Tr \left(\hat{U}(t)\right) \right) \right]$ for the matrix-valued GBM with non-commutative matrices.

For $q_z\ge q \ge 0$, taking the sum on $M$, we obtain the partition function of the odd number as follows:
\be
&&\Tr\left( e^{-\beta\hat{H}_{LMG}(n)  }\right)
\no\\
&=& \int Dw \int Du  \sum_{S=1/2,3/2,\cdots,n/2}^{}  {}_n C _{\frac{n}{2}-S} \frac{1+2S}{1+S+\frac{n}{2}} e^{A_2 S}\frac{\sinh(A_1(S+1/2))}{\sinh(A_1/2)} .
\label{partition-odd2}
\ee
We note that in Eq. (\ref{partition-odd2}), the sum variable takes $S=1/2,3/2,\cdots,n/2$. Thus, $S$ takes only half-integer values and the operation of taking the limit $n\to0$ is not justified.
Instead, it is possible to take the limit $n\to1$ (similarly, we can take the limit $n\to 2k+1$ for any natural number $k$).
 Then, the following identity is useful:
\be
\lim_{n\rightarrow1}    \frac{ \Tr\left( e^{-\beta\hat{H}_{LMG}(n)  }\right)- \Tr\left( e^{-\beta\hat{H}_{LMG}(1)  }\right)  }{n-1} 
&=&\lim_{n\rightarrow1} \mathbb{E} \left[   \frac{ \left( \Tr \left(\hat{U}(t)\right) \right)^{n}-  \Tr \left(\hat{U}(t)\right)  }{n-1} \right] 
\no\\
&=&\mathbb{E} \left[ \Tr \left(\hat{U}(t)\right)\log\left( \Tr \left(\hat{U}(t)\right) \right) \right]  .
\ee
Using this identity and by the same manner as before (see Appendix A for the derivation), we can obtain an analytical expression of $\mathbb{E} \left[ \Tr \left(\hat{U}(t)\right)\log\left( \Tr \left(\hat{U}(t)\right) \right) \right] $ for $q_z\ge q \ge 0$,
\be
&&\mathbb{E} \left[ \Tr \left(\hat{U}(\beta)\right)\log\left( \Tr \left(\hat{U}(\beta)\right) \right) \right] 
\no\\
&=& \int Dw \int Du\frac{e^{(-A_1-3A_2)/2}\left( (e^{2A_1}-e^{2A_2})\log(1+e^{-A_1+A_2}) +(-1+e^{2(A_1+A_2)}) \log(1+e^{A_1+A_2}) \right)}{2(-1+e^{A_1})}  , \label{Tr-LogTr}
\ee
and for $q\ge q_z \ge 0$,
\be
&&\mathbb{E} \left[ \Tr \left(\hat{U}(\beta)\right)\log\left( \Tr \left(\hat{U}(\beta)\right) \right) \right] 
\no\\
&=& \int Dw \int Du \Re\left(\frac{e^{(-A_3-3A_2)/2}\left( (e^{2A_3}-e^{2A_2})\log(1+e^{-A_3+A_2}) +(-1+e^{2(A_3+A_2)}) \log(1+e^{A_3+A_2}) \right)}{2(-1+e^{A_3})}  \right). \label{Tr-LogTr-2}
\ee

\begin{figure}[H]
\begin{tabular}{cc}
\begin{minipage}[t]{0.45\hsize}
\centering
\includegraphics[scale=0.53]{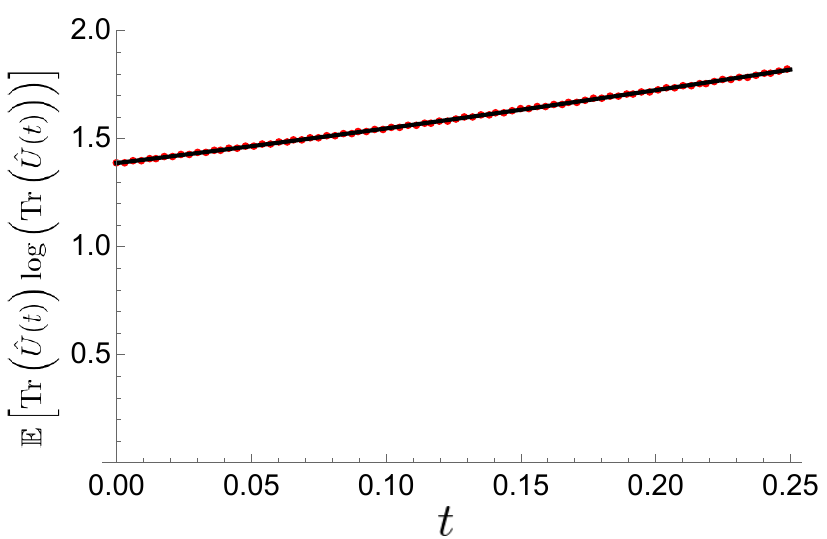}
\end{minipage} &
%%%%%%%%%%%%%%%%%%%%%%%%%%%%%%%%%%%%%%%%%%%%%%%%%%%%%%%%%%%%%%%%%
\begin{minipage}[t]{0.45\hsize}
\centering
\includegraphics[scale=0.53]{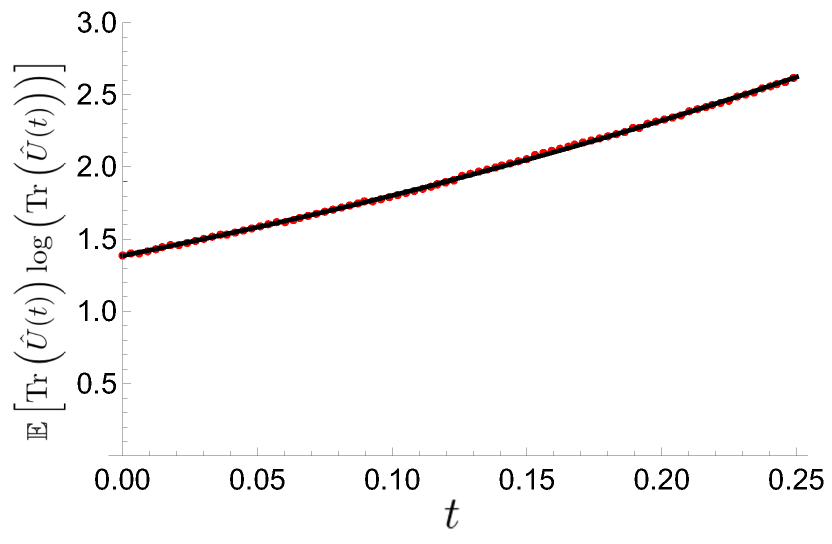}
\end{minipage}
\end{tabular}
\caption{Time dependence of $\mathbb{E} \left[ \Tr \left(\hat{U}(t)\right)\log\left( \Tr \left(\hat{U}(t)\right) \right) \right]$ for the matrix-valued GBM with non-commutative matrices. 
The horizontal axis denotes the time. The vertical axis denotes $\mathbb{E} \left[ \Tr \left(\hat{U}(t)\right)\log\left( \Tr \left(\hat{U}(t)\right) \right) \right]$.
Left panel: For $h=1/5$, $q=1/5$, and $q_z=1/2$, the thick black line and the red circles show Eq. (\ref{Tr-LogTr}) and the numerical simulation of Eq. (4), respectively.
Right pane: For $h=1/10$, $q=1$ and $q_z=1/5$, the thick black line and the red circles show Eq. (\ref{Tr-LogTr-2}) and the numerical simulation of Eq. (4), respectively.}
\label{fig2}
\end{figure}

In order to check our analytical expressions, we compared Eqs. (\ref{Tr-LogTr}) and (\ref{Tr-LogTr-2}) with the numerical simulation of Eq. (\ref{Ito}) in Fig. \ref{fig2}.
The numerical conditions are the same as in Fig. \ref{fig1}.
The numerical calculation shows that our analytical expression precisely coincides with the result of the numerical simulation of Eq. (\ref{Ito}).
Therefore, this verifies that we have obtained the exact solution of $\mathbb{E} \left[ \Tr \left(\hat{U}(\beta)\right)\log\left( \Tr \left(\hat{U}(\beta)\right) \right) \right]$.

%%%%%%%%%%%%%%%%%%%%%%%%%%%%%%%%%%%%%%%%%%%%%%%%%%%%%%%%%%%%%%%%%

%%%%%%%%%%%%%%%%%%%%%%%%%%%%%%%%%%%%%%%%%%%%%%%%%%%%%%%%%%%%%%%%%
\section{Conclusions}
We have obtained the exact solution of the free entropy in the matrix-valued GBM with non-commutative matrices.
The exact solution for a single trajectory has not been previously obtained for the matrix-valued GBM with non-commutative matrices. However, we demonstrate that the replica method enables us to obtain the exact solution of the free entropy in the matrix-valued GBM with non-commutative matrices through solving the partition function of the isotropic Lipkin-Meshkov-Glick model.
A similar analysis of the free entropy using the replica method has been performed for the KPZ equation~\cite{CDR,Dotsenko2}, and our result extends their results to a simple but nontrivial discrete system.

In the replica analysis, we had to treat the number of replicas separately for even and odd numbers.
This strange behavior does not arise in the conventional replica analysis of mean-field spin glass models~\cite{Nishimori} or the KPZ equation~\cite{CDR,Dotsenko2} and is due to the fact that the eigenvalues of the isotropic LMG model depend on the parity of the number of replicas in Eq. (\ref{eigenvalue-S}).

In this study, from the analogy of the free entropy in statistical mechanics, we have focused only on the logarithm of the trace of the time evolution operator.
It is an interesting problem to compute exact solutions for other quantities by the same method.

We note that for $q_x\neq q_y$, we have not succeeded in obtaining an exact solution of the free entropy. 
In this case, the eigenvalues of the Lipkin-Meshkov-Glick model have been exactly obtained using the Bethe ansatz~\cite{PD,PD2,MOPN}.
However, they have a very complex form, which makes it difficult to compute the partition function in Eq. (\ref{LMG-partition}).

%%%%%%%%%%%%%%%%%%%%%%%%%%%%%%%%%%%%%%%%%%%%%%%%%%%%%%%%%%%%%%%%%%
%\begin{acknowledgment}
The authors thank Tomoyuki Obuchi and Kentaro Ohki for useful discussions.
The present work was financially supported by JSPS KAKENHI Grant No. 18H03303, 19H01095, 19K23418, 21K13848 and the JST-CREST (No.JPMJCR1402) for Japan Science and Technology Agency.
%\end{acknowledgment}

%%%%%%%%%%%%%%%%%%%%%%%%%%%%%%%%%%%%%%%%%%%%%%%%%%%%%%%%%%%%%%%%%
\appendix
%%%%%%%%%%%%%%%%%%%%%%%%%%%%%%%%%%%%%%%%%%%%%%%%%%%%%%%%%%%%%%%%%
\section{Derivation of exact solution of $\mathbb{E} \left[ \Tr \left(\hat{U}(t)\right)\log\left( \Tr \left(\hat{U}(t)\right) \right) \right]$ } 
From Eq. (38), $\Tr\left( e^{-\beta\hat{H}_{LMG}(1)  }\right)  $ can be easily obtained as
\be
\Tr\left( e^{-\beta\hat{H}_{LMG}(1)  }\right)
&=& \int Dw \int Du   e^{A_2/2}\frac{\sinh(A_1 )}{\sinh(A_1/2)} .
\ee
On the other hand, we can represent $\Tr\left( e^{-\beta\hat{H}_{LMG}(n)  }\right)$ as
\be
&&\Tr\left( e^{-\beta\hat{H}_{LMG}(n)  }\right)
\no\\
&=& \int Dw \int Du \frac{1}{(-1+e^{A_1})(3+n)(5+n)}e^{(-3A_1+A_2)/2} 
\no\\
&&\left\{  4 e^{A_1}(5+n) {}_n C_{\frac{n-1}{2}} \left( -{}_2 F_{1} \left[1,\frac{1-n}{2},\frac{5+n}{2}, -e^{-A_1+A_2}\right]  +e^{2A_1}  {}_2 F_{1} \left[1,\frac{1-n}{2},\frac{5+n}{2}, -e^{A_1+A_2}\right]  \right) \right.
\no\\
&& \left.  -4e^{A_2} (3+n) {}_n C_{\frac{n-3}{2}} \left({}_2 F_{1}\left[2,\frac{3-n}{2},\frac{7+n}{2}, -e^{-A_1+A_2}\right]  - e^{4A_1}{}_2 F_{1} \left[2,\frac{3-n}{2},\frac{7+n}{2}, -e^{A_1+A_2}\right]  \right) \right\}.
\no\\
\ee
Then, using an analytic continuation, we can realize the $n\rightarrow 1$ limit as follows:
\be
&&\lim_{n\rightarrow1}    \frac{ \Tr\left( e^{-\beta\hat{H}_{LMG}(n)  }\right)- \Tr\left( e^{-\beta\hat{H}_{LMG}(1)  }\right)  }{n-1} 
\no\\
&=&\int Dw \int Du\frac{1}{4(-1+e^{A_1})}e^{(-A_1-3A_2)/2}
\no\\
&&   \left( -3 e^{2A_2}+3 e^{2(A_1+A_2)} +4(e^{2A_1}+e^{A_1 +A_2})\log(1+e^{-A_1+A_2}) -4(1+e^{A_1+A_2})\log(1+e^{A_1+A_2}) \right.
\no\\
&&\left. -2e^{2A_2}   {}_2 F_{1}^{(0,0,1,0)} \left[1,0,3, -e^{-A_1+A_2}\right] +2e^{2(A_1+A_2)}   {}_2 F_{1}^{(0,0,1,0)} \left[1,0,3, -e^{A_1+A_2}\right] \right.
\no\\
&& \left. 2e^{2A_2}   {}_2 F_{1}^{(0,1,0,0)} \left[1,0,3, -e^{-A_1+A_2}\right] -2e^{2(A_1+A_2)}   {}_2 F_{1}^{(0,1,0,0)} \left[1,0,3, -e^{A_1+A_2}\right] \right) .
\ee 
Furthermore, using the relations
\be
{}_2 F_{1}^{(0,0,1,0)} \left[1,0,3, y\right] &=&0,
\\
{}_2 F_{1}^{(0,1,0,0)} \left[1,0,3, y \right] &=&\frac{y(-2+3y)-2(-1+y)^2\log(1-y)}{2y^2},
\ee
we obtain an analytical expression of $\mathbb{E} \left[ \Tr \left(\hat{U}(t)\right)\log\left( \Tr \left(\hat{U}(t)\right) \right) \right]$ for $q_z\ge q \ge 0$,
\be
&&\mathbb{E} \left[ \Tr \left(\hat{U}(t)\right)\log\left( \Tr \left(\hat{U}(t)\right) \right) \right] 
\no\\
&=& \int Dw \int Du\frac{e^{(-A_1-3A_2)/2}\left( (e^{2A_1}-e^{2A_2})\log(1+e^{-A_1+A_2}) +(-1+e^{2(A_1+A_2)}) \log(1+e^{A_1+A_2}) \right)}{2(-1+e^{A_1})}  .\label{app-1}
\ee
For $q\ge q_z \ge 0$, an analytical expression is given by replacing $A_1$ with $A_3$ in Eq. (\ref{app-1}),
\be
&&\mathbb{E} \left[ \Tr \left(\hat{U}(t)\right)\log\left( \Tr \left(\hat{U}(t)\right) \right) \right] 
\no\\
&=& \int Dw \int Du \Re\left(\frac{e^{(-A_3-3A_2)/2}\left( (e^{2A_3}-e^{2A_2})\log(1+e^{-A_3+A_2}) +(-1+e^{2(A_3+A_2)}) \log(1+e^{A_3+A_2}) \right)}{2(-1+e^{A_3})}  \right).
\ee

%%%%%%%%%%%%%%%%%%%%%%%%%%%%%%%%%%%%%%%%%%%%%%%%%%%%%%%%%%%%%
%%%%%%%%%%%%%%%%%%%%%%%%%%%%%%%%%%%%%%%%%%%%%%%%%%%%%%%%%%%%%%%%%%%%%%%%%%%%

\end{document}